\documentclass[a4,10pt]{cernart}
\usepackage{graphicx}
\usepackage{fancyhdr}
\usepackage{enumerate}
\usepackage{indentfirst}
\usepackage{xspace}

\fancypagestyle{plain}{
 \fancyhf{}

}

\newcommand{\pionm}{\ensuremath{\pi^{-}}}
\newcommand{\pionp}{\ensuremath{\pi^{+}}}

\newcommand{\pip}{\ensuremath{\pi^{+}}\xspace}

\newcommand{\pim}{\ensuremath{\pi^{-}}\xspace}

\newcommand{\rad}{\ensuremath{\mbox{rad}}\xspace}

\newcommand{\GeVc}{\ensuremath{\mbox{GeV}/c}\xspace}
\newcommand{\MeVc}{\ensuremath{\mbox{MeV}/c}\xspace}

\newcommand{\mm}{\ensuremath{\mbox{mm}}\xspace}

\newcommand{\mrad}{\ensuremath{\mbox{mrad}}\xspace}

\newcommand{\ps}{\ensuremath{\mbox{ps}}\xspace}

\begin{document}
\begin{titlepage}
\docnum{CERN--PH--EP/2008--010}
\date{6 February 2008}

\vspace{2cm}

\title{\bf  {\Large Forward $\pi^\pm$ production in p--O$_2$ 
and p--N$_2$ interactions \\ at 12~GeV/$c$}}

\author{\large{HARP Collaboration}}
\maketitle

\vspace{2cm}

\begin{abstract}

Measurements of double-differential charged pion production 
cross-sections in interactions of 12~\GeVc protons on O$_2$ and N$_2$ 
thin targets are presented 
in the kinematic range
0.5~\GeVc  $\leq p_{\pi} <$ 8~\GeVc
and 50~mrad $\leq \theta_{\pi} <$ 250~mrad (in the laboratory frame)
and are compared with p--C results.  
For p--N$_2$ (p--O$_2$) interactions the analysis is performed using 
38576 (7522) reconstructed secondary pions. 
The analysis uses the beam instrumentation and the forward spectrometer
of the HARP experiment at CERN PS. 
The measured cross-sections have a direct impact on the precise
calculation of atmospheric neutrino fluxes and on the improved 
reliability of extensive air shower simulations by reducing 
the uncertainties of hadronic interaction models in the low energy range.  
In particular, the present results allow the common hypothesis that p--C
 data can be used to predict the p--N$_2$ and p--O$_2$ pion production
cross-sections to be tested.

\end{abstract} 

\vskip 2cm
\submitted{Submitted to Astroparticle Physics}
\end{titlepage}

\clearpage
\thispagestyle{plain}
\thispagestyle{plain}
\begin{center}
\vskip 2cm
{\small
M.G.~Catanesi, 
E.~Radicioni
\\ 
{\bf Sezione INFN, Bari, Italy} 
\\
R.~Edgecock, 
M.~Ellis$^{1}$          
\\
{\bf Rutherford Appleton Laboratory, Chilton, Didcot, UK} 
\\
C.~G\"{o}\ss ling 
\\
{\bf Institut f\"{u}r Physik, Universit\"{a}t Dortmund, Germany} 
\\
S.~Bunyatov, 
A.~Krasnoperov, 
B.~Popov$^2$, 
V.~Tereschenko 
\\
{\bf Joint Institute for Nuclear Research, JINR Dubna, Russia} 
\\
E.~Di~Capua, 
G.~Vidal--Sitjes$^{3}$  
\\
{\bf Universit\`{a} degli Studi e Sezione INFN, Ferrara, Italy}  
\\
A.~Artamonov$^4$,   
S.~Giani, 
S.~Gilardoni,       
P.~Gorbunov$^{4}$,  
A.~Grant,  
A.~Grossheim$^{6}$, 
A.~Ivanchenko$^{16}$,  
V.~Ivanchenko$^{7}$,  
A.~Kayis-Topaksu$^{8}$,
J.~Panman, 
I.~Papadopoulos,  
E.~Tcherniaev, 
I.~Tsukerman$^4$,   
C.~Wiebusch$^{9}$,    
P.~Zucchelli$^{5,10}$ 
\\
{\bf CERN, Geneva, Switzerland} 
\\
A.~Blondel, 
S.~Borghi$^{11}$,  
M.C.~Morone$^{12}$, 
G.~Prior$^{13}$,   
R.~Schroeter
\\
{\bf Section de Physique, Universit\'{e} de Gen\`{e}ve, Switzerland} 
\\
C.~Meurer
\\
{\bf Institut f\"{u}r Physik, Forschungszentrum Karlsruhe, Germany}
\\
\newcommand{\afkyot}{{19}\xspace}
U.~Gastaldi
\\
{\bf Laboratori Nazionali di Legnaro dell' INFN, Legnaro, Italy} 
\\
\newcommand{\aflanl}{{14}\xspace}
G.~B.~Mills$^{\aflanl}$  
\\
{\bf Los Alamos National Laboratory, Los Alamos, USA} %
\\
J.S.~Graulich$^{15}$, 
G.~Gr\'{e}goire 
\\
{\bf Institut de Physique Nucl\'{e}aire, UCL, Louvain-la-Neuve,
  Belgium} 
\\
M.~Bonesini,
F.~Ferri           
\\
{\bf Sezione INFN Milano Bicocca, Milano, Italy} 
\\
M.~Kirsanov
\\
{\bf Institute for Nuclear Research, Moscow, Russia} 
\\
A. Bagulya, 
V.~Grichine,  
N.~Polukhina
\\
{\bf P. N. Lebedev Institute of Physics (FIAN), Russian Academy of
Sciences, Moscow, Russia} 
\\
V.~Palladino
\\
{\bf Universit\`{a} ``Federico II'' e Sezione INFN, Napoli, Italy} 
\\
\newcommand{\afclmb}{{14}\xspace}
L.~Coney$^{\afclmb}$, 
D.~Schmitz$^{\afclmb}$
\\
{\bf Columbia University, New York, USA} %
\\
G.~Barr 
\\
{\bf Nuclear and Astrophysics Laboratory, University of Oxford, UK} 
\\
F.~Bobisut$^{a,b}$, 
D.~Gibin$^{a,b}$,
A.~Guglielmi$^{b}$, 
M.~Mezzetto$^{b}$
\\
{\bf Universit\`{a} degli Studi$^{a}$ e Sezione INFN$^{b}$, Padova, Italy} 
\\
J.~Dumarchez
\\
{\bf LPNHE, Universit\'{e}s de Paris VI et VII, Paris, France} 
\\
U.~Dore
\\
{\bf Universit\`{a} ``La Sapienza'' e Sezione INFN Roma I, Roma,
  Italy} 
\\
D.~Orestano$^{c,d}$, 
F.~Pastore$^{c,d}$, 
A.~Tonazzo$^{c,d}$, 
L.~Tortora$^{d}$
\\
{\bf Universit\`{a} degli Studi$^{c}$ e Sezione INFN$^{d}$ Roma III, Roma, Italy}
\\
C.~Booth, 
L.~Howlett
\\
{\bf Dept. of Physics, University of Sheffield, UK} 
\\
M.~Bogomilov, 
D.~Kolev, 
R.~Tsenov
\\
{\bf Faculty of Physics, St. Kliment Ohridski University, Sofia,
  Bulgaria} 
\\
S.~Piperov, 
P.~Temnikov
\\
{\bf Institute for Nuclear Research and Nuclear Energy, 
Academy of Sciences, Sofia, Bulgaria} 
\\
M.~Apollonio$^{17}$, 
P.~Chimenti,  
G.~Giannini
\\
{\bf Universit\`{a} degli Studi e Sezione INFN, Trieste, Italy} 
\\
J.~Burguet--Castell, 
A.~Cervera--Villanueva, 
J.J.~G\'{o}mez--Cadenas, 
J. Mart\'{i}n--Albo,
M.~Sorel
\\
{\bf  Instituto de F\'{i}sica Corpuscular, IFIC, CSIC and Universidad de Valencia,
Spain} 
}
\end{center}
\thispagestyle{plain}
\vfill
\rule{0.3\textwidth}{0.4mm}
\newline
$^{~1}${Now at FNAL, Batavia, Illinois, USA.}
\newline
$^{~2}${Also supported by LPNHE, Paris, France.}
\newline
%
$^{~3}${Now at Imperial College, University of London, UK.}
\newline
$^{~4}${ITEP, Moscow, Russian Federation.}
\newline
$^{~5}${Now at SpinX Technologies, Geneva, Switzerland.}
\newline
$^{~6}${Now at TRIUMF, Vancouver, Canada.}
\newline
$^{~7}${On leave of absence from Ecoanalitica, Moscow State University,
Moscow, Russia.}
\newline
$^{~8}${Now at \c{C}ukurova University, Adana, Turkey.}
\newline
$^{9}${Now at III Phys. Inst. B, RWTH Aachen, Aachen, Germany.}
\newline
$^{10}$On leave of absence from INFN, Sezione di Ferrara, Italy.
\newline
$^{11}${Now at CERN, Geneva, Switzerland.}
\newline
$^{12}${Now at University of Rome Tor Vergata, Italy.}
\newline
$^{13}${Now at Lawrence Berkeley National Laboratory, Berkeley, California, USA.}
\newline
$^{14}${MiniBooNE Collaboration.}
\newline
$^{15}${Now at Section de Physique, Universit\'{e} de Gen\`{e}ve, Switzerland, Switzerland.}
\newline
$^{16}${On leave from Novosibirsk State University, Novosibirsk, Russia.}
\newline
$^{17}${Now at Nuclear and Astrophysics Laboratory, Oxford University, UK.}
%

\clearpage

\section{Introduction}
\label{sec:intro}

The HARP experiment~\cite{ref:harp-prop} at the CERN PS
was designed to measure hadron yields from a large range
of nuclear targets and for incident particle momenta from 1.5~\GeVc to 15~\GeVc.
This corresponds to a proton momentum region of great interest for
neutrino beams and far from being covered by earlier dedicated hadroproduction
experiments~\cite{ref:na56,ref:atherton}.
The main motivations are the measurement of pion yields for a quantitative
design of the proton driver of a future neutrino factory~\cite{ref:nufact}, 
a substantial improvement in the calculation of the atmospheric neutrino
fluxes~\cite{ref:atm_nu_flux}
and the measurement of particle yields as input for the flux
calculation of accelerator neutrino experiments~\cite{ref:physrep},
such as K2K~\cite{ref:k2k,ref:k2kfinal},
MiniBooNE~\cite{ref:miniboone} and SciBooNE~\cite{ref:sciboone}.

Measurements 
of the double-differential cross-section
for $\pi^{\pm}$ production at large angles by
protons in the momentum range of 3~\GeVc--12.9~\GeVc impinging
on different thin 5\% nuclear interaction length ($\lambda_{\mathrm{I}}$) 
targets have been reported in~\cite{ref:harp:tantalum,
ref:harp:carboncoppertin,ref:harp:bealpb,ref:finalproton}. 
These measurements are of special interest for target 
materials used in conventional
accelerator neutrino beams and in neutrino factory designs.

The results on the forward production of $\pip$ in p--Al interactions at 12.9~\GeVc 
and p--Be interactions at 8.9~\GeVc, useful for the understanding of
the accelerator neutrino fluxes in the 
K2K, MiniBooNE and SciBooNE experiments, have been published in references
\cite{ref:alPaper,ref:bePaper}.

In this paper we address another of the  main motivations of the HARP
experiment: the measurement of the yields of positive and negative
pions relevant for a precise calculation of the atmospheric neutrino
fluxes and improved modeling of extensive air showers. 
We present measurements of the double-differential cross-section 
$
{{\mathrm{d}^2 \sigma^{\pi}}}/{{\mathrm{d}p\mathrm{d}\Omega }}
$
for positive and negative pion 
production (in the kinematic range of momentum
0.5~\GeVc  $\leq p_{\pi} <$ 8~\GeVc and angle
50~mrad $\leq \theta_{\pi} <$ 250~mrad in the laboratory frame) 
by protons 
of 12~\GeVc momentum impinging
on thin cryogenic N$_2$ and O$_2$  targets of  5.5\% and 7.5\%  
nuclear interaction length ($\lambda_{\mathrm{I}}$), respectively.
Results for the pion production on a thin carbon target
in almost the same kinematic region have already
been published in \cite{ref:carbonfw}.
Some of those results will
be shown again in this paper with a different binning for comparison
(see Appendix A).
These measurements are performed using the forward spectrometer of the
HARP detector.
Results on the measurement 
of the double-differential $\pi^{\pm}$ production
cross-section  in proton--carbon collisions 
obtained with the HARP large-angle spectrometer 
($100~\MeVc \leq p_{\pi} < 800~\MeVc$ 
and $0.35~\rad \le \theta_{\pi} <2.15~\rad$)
are presented in 
a separate article~\cite{ref:harp:carboncoppertin}. 

The existing world data for $\pi^{\pm}$ production on light targets 
at low beam momentum ($\leq$~25~{\GeVc}) are  mainly 
restricted to beryllium targets and with a limited phase space 
coverage~\cite{Baker61,Dekkers65,Allaby70,Cho71a,Eichten72,Antreasyan79}. 
The work of Eichten et al.~\cite{Eichten72} 
has become a widely used standard reference data set. 
In addition to these data, new results from the E910 Collaboration have
been recently published~\cite{ref:E910}.

Carbon is an isoscalar nucleus as 
nitrogen and oxygen, so the extrapolation to air is the most 
straightforward. 
%
Recently the p--C data at 158~\GeVc provided by the NA49 experiment 
at CERN SPS in a large acceptance range have become available~\cite{Alt:2006fr}. 
Relevant data are expected also from the MIPP experiment at Fermilab~\cite{MIPP}. 
We would like to mention that the NA61 experiment~\cite{NA61} took  first 
p--C data at 30~\GeVc in the autumn of 2007. The foreseen measurements 
of importance for astroparticle physics are studies of p--C interactions 
at incoming beam momenta 30~\GeVc, 40~\GeVc, 50~\GeVc and  $\pi^\pm$--C 
interactions at 158~\GeVc and 350~\GeVc.

It is more difficult for experiments to study p--O$_2$ and p--N$_2$ reactions 
because cryogenic targets are more complicated to handle.
The results presented in this paper are the first for this
type of targets in this energy range.

\subsection{Experimental apparatus}
\label{subsec:harp_det}

\begin{figure}[tbh]
\centering
\includegraphics[width=0.8\textwidth]{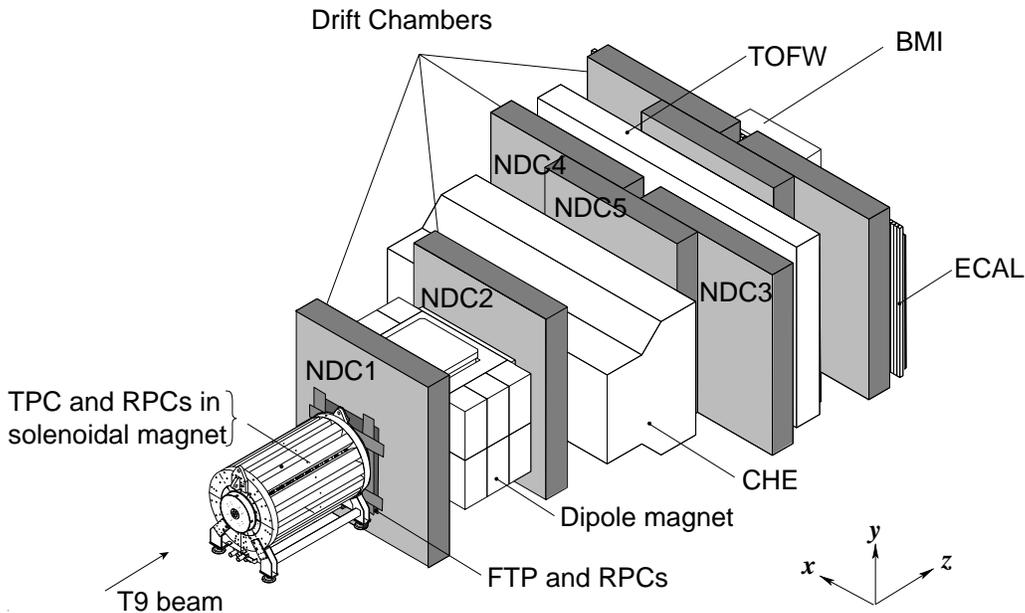} 
\caption{\label{fig:harp} 
Schematic layout of the HARP detector.
The convention for the coordinate system is shown in the lower-right
corner.
}
\end{figure}

 The HARP experiment~\cite{ref:harp-prop,ref:harpTech}
 makes use of a large-acceptance spectrometer consisting of a
 forward and large-angle detection system.
 The HARP detector is shown in Fig.~\ref{fig:harp}.
 A detailed
 description of the experimental apparatus can be found in Ref.~\cite{ref:harpTech}.
 The forward spectrometer -- 
 based on five modules of large area drift chambers
 (NDC1-5)~\cite{ref:NOMAD_NIM_DC} and a dipole magnet
 complemented by a set of detectors for particle identification (PID): 
 a time-of-flight wall (TOFW)~\cite{ref:tofPaper}, a large Cherenkov detector (CHE) 
 and an electromagnetic calorimeter (ECAL) --
 covers polar angles up to 250~mrad. 
 The muon contamination of the beam is measured with a muon identifier 
 consisting of thick iron absorbers and scintillation counters.
 The large-angle spectrometer -- based on a Time Projection Chamber (TPC) 
 and Resistive Plate Chambers (RPCs)
 located inside a solenoidal magnet --
 has a large acceptance in the momentum
 and angular range for the pions relevant to the production of the
 muons in a neutrino factory. 
 For the analysis described here  only  the forward spectrometer and
 the beam instrumentation are used.

The HARP experiment, located in the T9 beam of the CERN PS, took data in 2001
and 2002.
The momentum definition of the T9 beam 
is known with a precision of the order of 1\%~\cite{ref:t9}. 

The target is placed inside the inner field cage (IFC) of the TPC,
in an assembly that can be moved in and out of the solenoid magnet.
In the cryogenic target setup used for N$_2$ and O$_2$, 
the gas was liquefied by thermal contact with a bath of helium,
compressed to 20 bar and then refrigerated to 13 K
by adiabatic expansion.
The refrigerator system was housed inside a vacuum cryostat (typically 
$2 \times 10^{-9}$ bar) ending in 
a vacuum tube containing the target. The target arm 
had 6 cm diameter, 250 $\mu$m thick mylar beam entrance and exit windows, 
and the space separating it from the IFC 
was flushed with nitrogen gas to avoid condensation.
The target used for the measurements presented here consisted of a 
6 cm long, 3 cm of diameter and 125 $\mu$m thick mylar 
cylinder and a curved downstream nose, 
for an actual target volume of about 75 cm$^3$. 

The thickness of the target is equivalent to about 5.5\% $\lambda_I$
(4.84~{g/cm$^{2}$}) for N$_2$ and 
7.5\% $\lambda_I$ (6.85~{g/cm$^{2}$}) for O$_2$.

The cooling causes a contraction of the target, which induces an uncertainty
on its thickness of the order of 1\%. This is taken into account in
the uncertainty on the number of target nuclei in section~\ref{errorest}.

A set of four multi-wire
proportional chambers (MWPCs) measures the position and direction of
the incoming beam particles with an accuracy of $\approx$1~\mm in
position and $\approx$0.2~\mrad in angle per projection.
A beam time-of-flight system (BTOF)
measures the time difference of particles over a $21.4$~m path-length. 
It is made of two
identical scintillation hodoscopes, TOFA and TOFB (originally built
for the NA52 experiment~\cite{ref:NA52}),
which, together with a small target-defining trigger counter (TDS,
also used for the trigger), provide particle
identification at low energies. This provides separation of pions, kaons
and protons up to 5~\GeVc and determines the initial time at the
interaction vertex ($t_0$). 
The timing resolution of the combined BTOF system is about 70~\ps.
A system of two N$_2$-filled Cherenkov detectors (BCA and BCB) is
used to tag electrons at low energies and pions at higher energies. 
The electron and pion tagging efficiency is found to be close to
100\%.
At the beam energy used for this analysis the Cherenkov counters select
all particles lighter than protons, while the BTOF is used to reject ions. 
A set of trigger detectors completes the beam instrumentation.

The selection of beam protons is performed using the same criteria as described
in~\cite{ref:alPaper}.
A downstream trigger in the forward scintillator trigger plane (FTP) 
was required to record the event,
accepting only tracks with a trajectory outside the central hole
(60~mm) which allows beam particles to pass. 

The length of the accelerator spill is 400~ms with a typical intensity
of 15\,000 beam particles per spill.
The average number of events recorded by the data acquisition ranges
from 300 to 350 per spill.

The absolute normalization of the number of incident protons was
performed using `incident-proton' triggers. 
These are triggers where the same selection on the beam particle was
applied but no selection on the interaction was performed.
The rate of this trigger was down-scaled by a factor 64.

\section{Data Analysis}
\subsection{Event and particle selection}

A detailed description of the experimental techniques 
used
for data analysis in the HARP forward spectrometer 
can be found in Ref.~\cite{ref:alPaper,ref:pidPaper}.

With respect to our first paper on pion production in
p--Al interactions~\cite{ref:alPaper}, a number of 
improvements to the analysis techniques 
and detector simulation have been made. 
The most important improvements introduced in this analysis 
compared with the one presented in Ref.~\cite{ref:alPaper} are:
\begin{itemize}
\item An increase 
of the track reconstruction efficiency;
\item Better understanding of the momentum scale and resolution of the detector, 
based on data, which was then used to tune the simulation;
\item New 
particle identification hit selection algorithms both in the 
TOFW and in the 
CHE resulting in much reduced background and negligible efficiency
      losses.  
 In the kinematic
range of the current analysis the pion identification efficiency is
about 98\%, while the background from mis-identified protons is well
below 1\%; 
\item Significant increases in Monte Carlo production  
have also reduced uncertainties from Monte Carlo statistics 
and allowed studies which have reduced certain systematics.
\end{itemize}

Further details of these improved analysis techniques 
can be found in~\cite{ref:bePaper,ref:carbonfw}.
For the cryogenic targets, dedicated, high statistics Monte Carlo data
were produced using an accurate description of the target geometry.

At the first stage of the analysis a  beam particle type is selected
using the beam time of flight system (TOF-A, TOF-B) and the Cherenkov
counters (BCA, BCB) as described in section~\ref{subsec:harp_det}.
A value of the pulseheight consistent with the pedestal in both beam
Cherenkov detectors rejects electrons, pions and kaons.
The beam TOF system is used to reject ions, such as deuterons, but at
12~\GeVc is not used to separate protons from pions.
However, we require time measurements in TOF-A, TOF-B and/or TDS to be
present which are needed for calculating 
the arrival time of the beam proton at the target. 


The purity of the proton beam is better than 99\%, with the
main background formed by kaons estimated to be 0.5\%.
This contamination is neglected in the analysis.

\begin{table}[tbp!] 
\caption{Total number of events and selected pions used in the carbon,
  nitrogen and oxygen 
  thin target analysis at 12 GeV/c, and the number of
  protons on target as calculated from the pre-scaled trigger count.} 
\label{tab:events}
{\small
\begin{center}
\begin{tabular}{lrrr} \hline
\bf{Data set}            &  C       & N$_2$   & O$_2$           \\ \hline
    Total DAQ events     &  1062429 & 1375780 & 246153          \\
 Acc. beam protons forward  &  4375230 & 5250240 & 840256         \\
   interactions             &          &         &     \\
  \bf{$\pi^-$ selected with PID} &  8179 &  14828 & 2817            \\
 \bf{$\pi^+$ selected with PID} &  13530 & 23748 & 4705              \\
                                   \hline
\end{tabular}
\end{center}
}
\end{table}

Secondary track selection criteria,
described in~\cite{ref:carbonfw},  are optimized to ensure the quality
of momentum reconstruction and a clean time-of-flight measurement
while maintaining a high reconstruction efficiency.

The background induced by
interactions of beam particles in the materials outside the target
is measured  by taking data without a
target in the target holder (``empty target data'').  
These data are  subject to the same  event and track
selection criteria as the standard  data sets. 

To take into account this background the number of particles of the
observed 
type (\pionp, \pionm) in the ``empty target data''
are subtracted bin-by-bin (momentum and angular bins) from the number
of particles of the same type. The uncertainty
induced by
this method is discussed in section~\ref{errorest} and
labeled ``empty target subtraction''. 
The event statistics is summarised in Table~\ref{tab:events}. 

\subsection{Cross-section calculation}

The cross-section is calculated as follows
\begin{eqnarray}
\frac{d^2 \sigma^{\alpha}}{dp d\Omega}(p_i,\theta_j) & = & 
\frac{A}{N_A \rho t} \cdot \frac{1}{N_{\rm pot}} \cdot \frac{1}{\Delta p_i \Delta \Omega_j} \cdot 
\sum_{p'_i,\theta'_j,\alpha'} \mathcal{M}^{\rm cor}_{p_i\theta_j\alpha p'_i\theta'_j\alpha'} \cdot 
N^{\alpha'}(p'_i,\theta'_j)\hspace{0.1cm},
\end{eqnarray} 
where 
\begin{itemize}
\item $\frac{d^2 \sigma^{\alpha}}{dp d\Omega}(p_i,\theta_j)$ is the
  cross-section in mb/(\GeVc sr) for the particle type $\alpha$ (p,
  \pionp or \pionm) for each true momentum and angle bin ($p_i,\theta_j$)
  covered in this analysis;
\item $N^{\alpha'}(p'_i,\theta'_j)$  is the number of particles of
  type $\alpha$ in bins of reconstructed momentum $p'_i$ and angle
  $\theta_j'$ in the raw data;
\item $\mathcal{M}^{\rm cor}_{p\theta\alpha p'\theta'\alpha'}$ is the
  correction matrix which accounts for efficiency and resolution of
  the detector;
\item $\frac{A}{N_A \rho t}$, $\frac{1}{N_{\rm pot}}$ and
  $\frac{1}{\Delta p_i \Delta \Omega_j}$ are normalization factors,
  namely:
\subitem $\frac{N_A \rho t}{A}$ is the number of target nuclei per unit area 
\footnote{$A$ - atomic  mass, $N_A$ - Avogadro number, $\rho$ - target
  density and $t$ - target thickness};
\subitem $N_{\rm pot}$ is the number of incident beam particles on
  target (particles on target);
\subitem $\Delta p_i $ and $\Delta \Omega_j $ are the bin sizes in
  momentum and solid angle, respectively 
\footnote{$\Delta p_i = p^{\rm max}_i-p^{\rm min}_i$,\hspace{0.2cm}
  $\Delta \Omega_j = 2 \pi (\cos(\theta^{\rm min}_j)- 
  \cos(\theta^{\rm max}_j))$}.
\end{itemize}
We do not make a correction for the attenuation
of the proton beam in the target, so that strictly speaking the
cross-sections are valid for a $\lambda_{\mathrm{I}}=5.5\%$ (7.5\%) 
N$_2$ (O$_2$) target.

The  calculation of the
correction matrix $M^{\rm cor}_{p_i\theta_j\alpha
  p'_i\theta'_j\alpha'}$ is 
a rather difficult task.
Various techniques are
described in the literature to obtain this matrix. As 
discussed 
in Ref.~\cite{ref:alPaper} for the p-Al analysis of HARP data at 12.9~{\GeVc}, two
complementary analyses have been performed to cross-check internal
consistency and possible biases in the respective procedures.
A comparison of both analyses shows that the results are consistent
within the overall systematic error~\cite{ref:alPaper}.

In the first method -- called ``Atlantic'' in~\cite{ref:alPaper} -- 
the correction matrix $M^{\rm
  cor}_{p_i\theta_j\alpha p'_i\theta'_j\alpha'}$ is decomposed into
distinct independent contributions, which are computed mostly using
the data themselves.
The second method -- called ``UFO'' in~\cite{ref:alPaper} -- 
is the unfolding method introduced 
by D'Agostini~\cite{ref:DAgostini}. 
It is based on the Bayesian unfolding technique.
In this case a simultaneous (three dimensional) unfolding of
momentum $p$, angle $\theta$ and particle type $\alpha$ is
performed. The correction matrix is computed using a Monte Carlo
simulation. This method has been used in the recent HARP 
publications~\cite{ref:harp:tantalum,ref:harp:carboncoppertin,ref:harp:bealpb} 
and it is also applied in the analysis described here 
(see~\cite{ref:carbonfw,ref:christine_phd} for additional information).

The Monte Carlo simulation of the HARP setup is based on 
GEANT4~\cite{ref:geant4}. 
The detector
materials are accurately 
described
in this simulation as well as the
relevant features of the detector response and the digitization
process. All relevant physics processes are considered, including
multiple scattering, energy loss, absorption and
re-interactions. 
The simulation is independent of the beam particle type
because it only generates for each event
exactly one secondary particle of a specific particle type inside the
target material and propagates it through the 
complete detector. 
A small difference (at the few percent level) is observed between the
efficiency calculated for 
events simulated with the single-particle Monte Carlo and with a
simulation using a multi-particle hadron-production model.
A similar difference is seen between the single-particle Monte Carlo and
the efficiencies measured directly from the data.
A momentum-dependent correction factor determined using the efficiency
measured with the  data is applied to take this into account. 
The track reconstruction used in this analysis and the simulation are
identical to the ones used for the \pionp\ production in p-Be
collisions~\cite{ref:bePaper}. 
A detailed description of the corrections and their magnitude can be
found there. 

The reconstruction efficiency (inside the geometrical acceptance) is
larger than 95\% above 1.5~\GeVc and drops to 80\% at 0.5~\GeVc. 
The requirement of a match with a TOFW hit has an efficiency between
90\% and 95\% independent of momentum.
The electron veto rejects about 1\% of the pions and protons below
3~\GeVc with a remaining background of less than 0.5\%.
Below Cherenkov threshold the TOFW separates pions and protons with
negligible background and an efficiency of $\approx$98\% for pions.
Above Cherenkov threshold the efficiency for pions is greater than 99\%
with only 1.5\% of the protons mis-identified as a pion.
The kaon background in the pion spectra is smaller than 1\%.

The absorption and decay of particles is simulated by the Monte Carlo.
The generated single particle can re-interact and produce background
particles by hadronic or electromagnetic processes, thus giving rise to
tracks in the dipole spectrometer.
In such cases also the additional measurements are entered into the
migration matrix thereby taking into account the combined effect of the
generated particle and any secondaries it creates.
The absorption correction is on average 20\%, approximately independent
of momentum.
Uncertainties in the absorption of secondaries in the dipole
spectrometer material are taken into account by
a variation of 10\% of this effect in the simulation. 
The effect of pion decay is treated in the same way as the absorption
and is 20\% at 500~\MeVc and negligible at 3~\GeVc. 

The uncertainty in the production of background due to tertiary
particles is larger. 
The average correction is $\approx$10\% and up to 20\% at
1~\GeVc. 
The correction includes reinteractions in the detector material as well
as a small component coming from reinteractions in the target.
The validity of the generators used in the simulation was checked by an
analysis of HARP data with incoming protons, and charged pions on
aluminium and carbon targets at lower momenta (3~\GeVc and 5~\GeVc).
A 30\% variation of the secondary production was applied.
The average empty-target subtraction amounts to $\approx$20\%.


Owing to the redundancy of the tracking system downstream of the
target the detection efficiency is very robust under the usual
variations of the detector performance during the long data taking
periods. 
Since the momentum is reconstructed without making use of the upstream
drift chamber module (which is more sensitive in its performance to the beam
intensity) the reconstruction efficiency is uniquely determined by the
downstream system.
No variation of the overall efficiency has been observed.
The performance of the TOFW and CHE system have been monitored to be
constant for the data taking periods used in this analysis.
The calibration of the detectors was performed on a day-by-day basis.

\subsection{Error estimation}
\label{errorest}


The total statistical error of the corrected data is composed of the
statistical error of the raw data and of the statistical error
of the unfolding procedure, as the unfolding matrix is obtained
from the data themselves, thus contributing also to the statistical
error. The statistical error provided by the unfolding program is
equivalent to the propagated statistical error of the raw data. In
order to calculate the statistical error of the unfolding procedure a
separate analysis is applied,
as described in~\cite{ref:carbonfw,ref:grossheim}. 
Its conclusion is that the statistical error provided by the unfolding
procedure has to be multiplied globally by a factor of 2, which is done
for the analyses described here.
This factor is somewhat dependent on the shape of the distributions.
For example a value 1.7 was found for the analysis reported in
Ref.~\cite{ref:harp:tantalum}. 

Different types of sources 
induce
systematic errors for the analysis
described here: 
track yield corrections ($\sim 5 \%$), particle identification ($\sim 0.1 \%$),
momentum and angular reconstruction ($\sim 0.5 \%$)~\footnote{
The quoted error in parenthesis refers to fractional error of the integrated cross-section
in the kinematic range covered by the HARP experiment}.
The strategy to calculate these systematic errors and the different
methods used for their evaluation are described in~\cite{ref:carbonfw}.
An additional source of error is due to misidentified secondary kaons, 
which are not considered in the particle identification method used for
this analysis and are subtracted on the
basis of a Monte Carlo simulation, as in~\cite{ref:carbonfw}. 
No explicit correction is made for pions coming from decays of other 
particles created in the target, as they give a very small contribution
according to the selection criteria applied in the analysis.

As a result of these systematic error studies, each error source 
can be represented by a covariance matrix. The
sum of these matrices describes the total systematic error,
as explained in~\cite{ref:carbonfw}. 

On average the total integrated systematic error is around $5-6\%$,
with a differential bin to bin systematic error of the order of
$10-11 \%$, to be compared with a statistical integrated (bin-to-bin
differential) error of $\sim 2-3 \%$ ($\sim 10-13 \%$).
Systematic and statistical errors are roughly of the same order. 
  
The overall normalization of the results 
is calculated relative to the number of incident beam
particles accepted by the selection. 
The uncertainty is 2\% for incident protons.
The contraction of the target with cooling induces an additional
systematic error of 1\% on the N$_2$ and O$_2$ data.

\section{Results}
\label{sec:results}

\begin{figure}[tb]
\centering
\includegraphics[width=.49\textwidth]{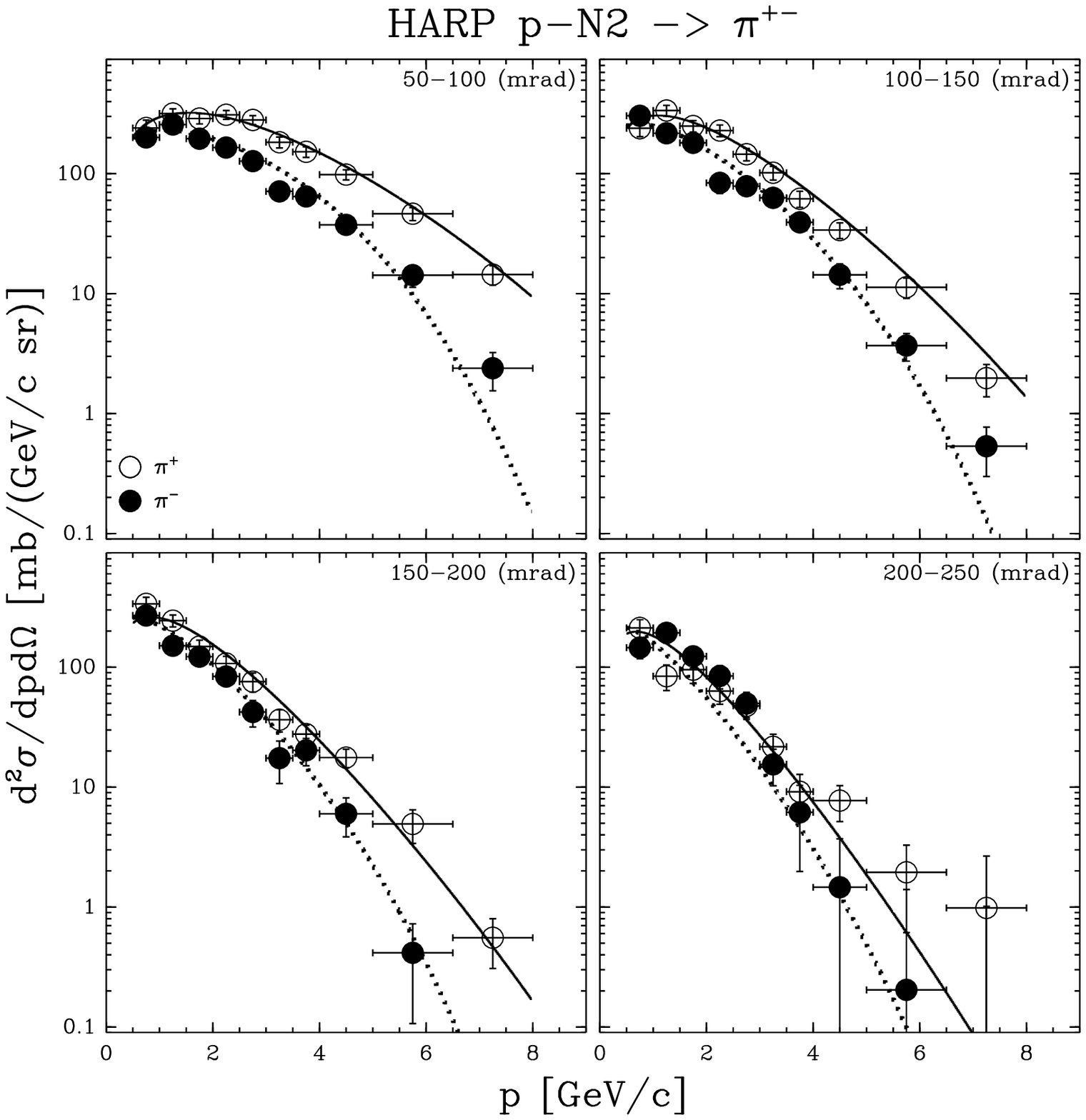}
\includegraphics[width=.49\textwidth]{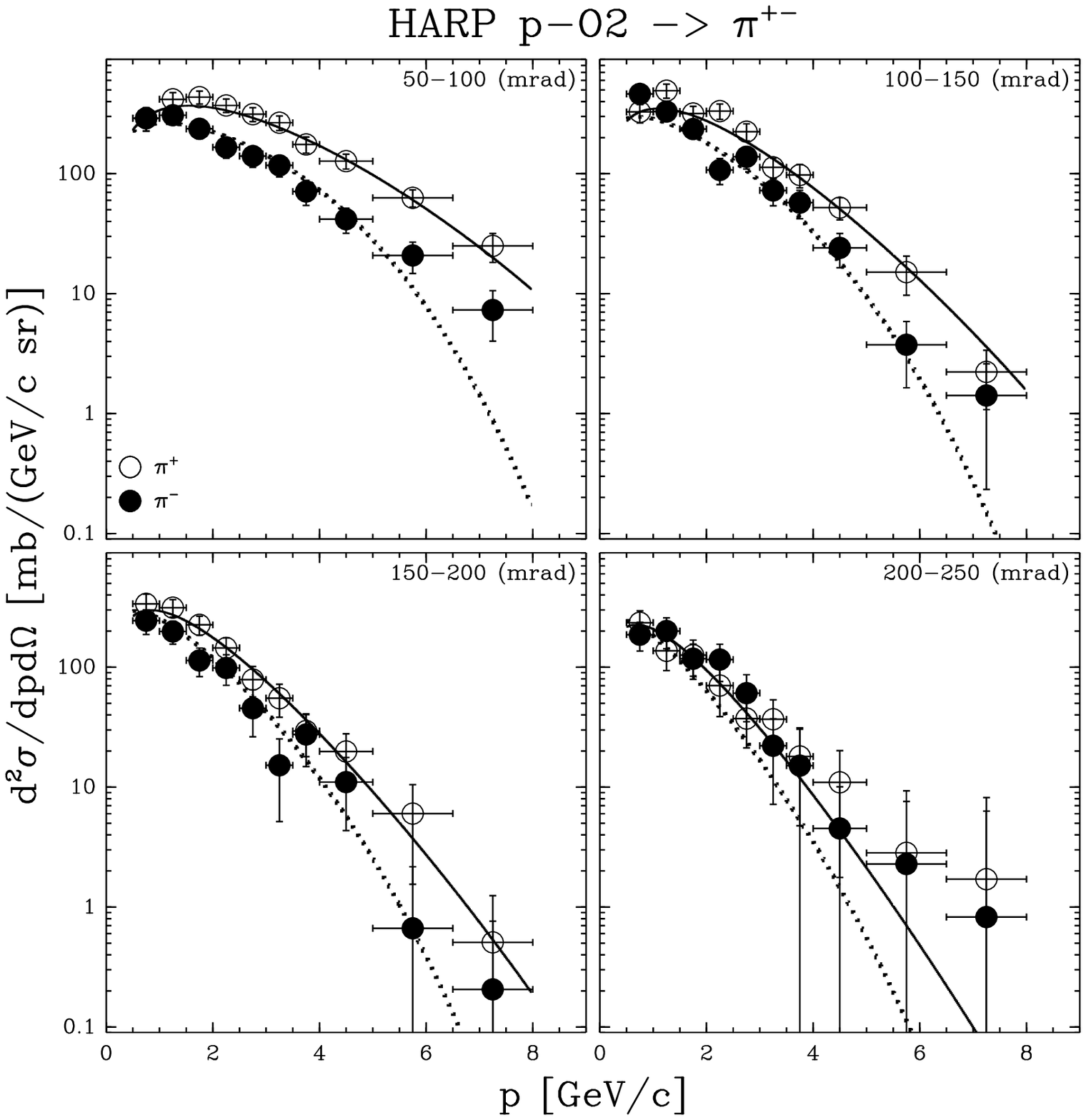}
\caption{
Measurement of the double-differential production cross-section of positive (open circles) and negative (filled circles) pions from 12~\GeVc protons on N$_2$ (left)
and O$_2$ (right) as a function of pion momentum, $p$, 
in bins of pion angle, $\theta$, in the laboratory frame. 
The curves show the Sanford-Wang parametrization with the parameters 
given in Ref.~\cite{ref:carbonfw} 
(solid line for $\pip$ and dashed line for $\pim$), 
computed for the central value of each angular bin.
In the top right corner of each plot the 
covered angular range is shown in mrad.}
\label{fig:san-wang}
\end{figure}

In Figure~\ref{fig:san-wang}, 
the measured $\pi^+$ and $\pi^-$ spectra in p--N$_2$ and p--O$_2$ interactions
at 12~\GeVc are compared to 
an empirical prametrization, developed by Sanford and Wang~\cite{SanfordWang1967}
to describe the production cross-sections of mesons in proton-nucleus
interactions.
The parameters fitted to our p--C data at 12~\GeVc in~\cite{ref:carbonfw} have been used 
and only a constant overall rescaling factor 
accounting for the target atomic mass has 
been applied.
One can observe that the shape and normalization obtained using the
carbon data predict quite well the nitrogen and oxygen data.  
This point will be made more clear when the N$_2$/C and O$_2$/C ratios are taken.
The shapes of the momentum spectra 
are similar for secondary \pionp\
and \pionm\ ,
as well as for different data sets, where only 
a different normalisation factor can be noticed because of the 
different nuclear masses of the target nuclei. 
The conclusions drawn in~\cite{ref:carbonfw} appear to be confirmed for the 
data sets presented here: the parametrization provides an approximate
description of the main features, but is not able to describe the data well
in some regions of kinematic space, particularly at high momenta and at
large angles.

The central values
and square-root of the diagonal elements of the covariance matrix are listed in
Tables~\ref{tab:xsec_results_pN2} and~\ref{tab:xsec_results_pO2}. 
\begin{table}[!h]
 \small{ 
 \caption{\label{tab:xsec_results_pN2}
    HARP results for the double-differential $\pi^+$ and $\pi^-$ production
    cross-section in the laboratory system,
    $d^2\sigma^{\pi}/(dpd\Omega)$, for p--N$_2$ interactions at 12~\GeVc. 
    Each row refers to a
    different $(p_{\hbox{\small min}} \le p<p_{\hbox{\small max}},
    \theta_{\hbox{\small min}} \le \theta<\theta_{\hbox{\small max}})$ bin,
    where $p$ and $\theta$ are the pion momentum and polar angle, respectively.
    The central value as well as the square-root of the diagonal elements
    of the covariance matrix are given.}
  \centerline{
    \begin{tabular}{|c|c|c|c|rcr|rcr|} \hline
$\theta_{\hbox{\small min}}$ &
$\theta_{\hbox{\small max}}$ &
$p_{\hbox{\small min}}$ &
$p_{\hbox{\small max}}$ &
\multicolumn{3}{c|}{$d^2\sigma^{\pi^+}/(dpd\Omega)$} &
\multicolumn{3}{c|}{$d^2\sigma^{\pi^-}/(dpd\Omega)$} 
\\
(rad) & (rad) & (GeV/c) & (GeV/c) &
\multicolumn{3}{c|}{(mb/(GeV/c sr))} &
\multicolumn{3}{c|}{(mb/(GeV/c sr))}
\\ \hline
 0.05 & 0.10 & 0.50 & 1.00& 240.5 &    $\pm$ &39.9 & 200.7 &  $\pm$ &33.7\\ 
      &      & 1.00 & 1.50& 318.1 &    $\pm$ &29.0 & 256.9 &  $\pm$ &27.9 \\ 
      &      & 1.50 & 2.00& 288.2 &    $\pm$ &28.1 & 195.9 &  $\pm$ &23.2 \\ 
      &      & 2.00 & 2.50& 310.5 &    $\pm$ &26.3 & 165.3 &  $\pm$ &18.0 \\ 
      &      & 2.50 & 3.00& 280.6 &    $\pm$ &23.9 & 127.4 &  $\pm$ &14.8 \\ 
      &      & 3.00 & 3.50& 183.0 &    $\pm$ &18.9 & 71.2  &  $\pm$ &10.1 \\ 
      &      & 3.50 & 4.00& 152.3 &    $\pm$ &15.6 & 64.7  &  $\pm$ &8.8 \\ 
      &      & 4.00 & 5.00&  98.3 &     $\pm$ & 9.5 & 37.5  &  $\pm$ &5.6 \\ 
      &      & 5.00 & 6.50&  46.4 &     $\pm$ &5.8 & 14.3  &  $\pm$ &3.0 \\ 
      &      & 6.50 & 8.00&  14.4 &     $\pm$ &2.7 & 2.4   &  $\pm$ &0.8 \\ 
 0.10 & 0.15 & 0.50 & 1.00& 239.5 &    $\pm$ &35.5 & 304.3 &  $\pm$ &42.8 \\ 
      &      & 1.00 & 1.50& 336.9 &    $\pm$ &34.7 & 217.3 &  $\pm$ &24.5 \\ 
      &      & 1.50 & 2.00& 250.0 &    $\pm$ &27.2 & 180.8 &  $\pm$ &20.9 \\ 
      &      & 2.00 & 2.50& 229.3 &    $\pm$ &25.6 & 83.9  &  $\pm$ &15.4 \\ 
      &      & 2.50 & 3.00& 145.3 &    $\pm$ &17.5 & 78.7  &  $\pm$ &11.9 \\ 
      &      & 3.00 & 3.50& 101.8 &    $\pm$ &13.4 & 63.0  &  $\pm$ &8.6  \\ 
      &      & 3.50 & 4.00&  61.9 &     $\pm$ &9.6 & 39.3  &  $\pm$ &6.5 \\ 
      &      & 4.00 & 5.00&  33.9 &     $\pm$ &5.1 & 14.3  &  $\pm$ &3.4 \\ 
      &      & 5.00 & 6.50&  11.3 &     $\pm$ &2.2 & 3.7   &  $\pm$ &1.0 \\ 
      &      & 6.50 & 8.00&   2.0 &     $\pm$ &0.6 & 0.5   &  $\pm$ &0.2 \\ 
 0.15 & 0.20 & 0.50 & 1.00& 337.6 &    $\pm$ &45.5 & 269.8 &  $\pm$ &40.0 \\ 
      &      & 1.00 & 1.50& 244.4 &    $\pm$ &27.8 & 151.1 &  $\pm$ &21.8 \\ 
      &      & 1.50 & 2.00& 148.6 &    $\pm$ &19.3 & 122.4 &  $\pm$ &18.7 \\ 
      &      & 2.00 & 2.50& 107.4 &    $\pm$ &15.8 & 83.7 &  $\pm$ &14.7 \\ 
      &      & 2.50 & 3.00&  75.8 &    $\pm$ &13.6 & 42.3 &  $\pm$ &10.5 \\ 
      &      & 3.00 & 3.50&  36.5 &     $\pm$ &7.7 & 17.5 &  $\pm$ &6.7 \\ 
      &      & 3.50 & 4.00&  27.7 &     $\pm$ &5.2 & 20.3 &  $\pm$ &5.2 \\ 
      &      & 4.00 & 5.00&  17.6 &     $\pm$ &3.1 & 6.0 &  $\pm$ &2.1 \\ 
      &      & 5.00 & 6.50&   4.9 &     $\pm$ &1.5 & 0.4 &  $\pm$ &0.3 \\ 
      &      & 6.50 & 8.00&   0.6 &     $\pm$ &0.2 & & - &   \\ 
 0.20 & 0.25 & 0.50 & 1.00& 212.6 &    $\pm$ &36.0 & 145.5 &  $\pm$ &28.0\\ 
      &      & 1.00 & 1.50&  84.1 &    $\pm$ &20.3 & 193.4 &  $\pm$ &34.2 \\ 
      &      & 1.50 & 2.00&  95.4 &    $\pm$ &22.3 & 123.2 &  $\pm$ &25.2 \\ 
      &      & 2.00 & 2.50&  63.2 &    $\pm$ &14.2 & 84.4 &  $\pm$ &18.5 \\ 
      &      & 2.50 & 3.00&  47.6 &    $\pm$ &10.9 & 49.8 &  $\pm$ &11.8 \\ 
      &      & 3.00 & 3.50&  21.7 &     $\pm$ &5.9 & 15.5 &  $\pm$ &5.2 \\ 
      &      & 3.50 & 4.00&   9.2 &     $\pm$ &3.6 & 6.2 &  $\pm$ &4.2 \\ 
      &      & 4.00 & 5.00&   7.7 &     $\pm$ &2.6 & 1.5 &  $\pm$ &2.2 \\ 
      &      & 5.00 & 6.50&   1.9 &     $\pm$ &1.3 & &  - & \\ 
      &      & 6.50 & 8.00&    &     -  & &  &  - & \\ 
\hline
\end{tabular}

  }
  }
\end{table}
\begin{table}[!h]
  \small{
  \caption{\label{tab:xsec_results_pO2}
    HARP results for the double-differential $\pi^+$ and $\pi^-$ production
    cross-section in the laboratory system,
    $d^2\sigma^{\pi}/(dpd\Omega)$, for p--O$_2$ interactions at 12~\GeVc. 
    Each row refers to a
    different $(p_{\hbox{\small min}} \le p<p_{\hbox{\small max}},
    \theta_{\hbox{\small min}} \le \theta<\theta_{\hbox{\small max}})$ bin,
    where $p$ and $\theta$ are the pion momentum and polar angle, respectively.
    The central value as well as the square-root of the diagonal elements
    of the covariance matrix are given.}
  \centerline{
    \begin{tabular}{|c|c|c|c|rcr|rcr|} \hline
$\theta_{\hbox{\small min}}$ &
$\theta_{\hbox{\small max}}$ &
$p_{\hbox{\small min}}$ &
$p_{\hbox{\small max}}$ &
\multicolumn{3}{c|}{$d^2\sigma^{\pi^+}/(dpd\Omega)$} &
\multicolumn{3}{c|}{$d^2\sigma^{\pi^-}/(dpd\Omega)$} 
\\
(rad) & (rad) & (GeV/c) & (GeV/c) &
\multicolumn{3}{c|}{(mb/(GeV/c sr))} &
\multicolumn{3}{c|}{(mb/(GeV/c sr))}
\\ \hline

 0.05 & 0.10 & 0.50 & 1.00& 290.3 &   $\pm$ & 63.9 & 290.6 &  $\pm$ & 63.3 \\ 
      &      & 1.00 & 1.50& 417.5 &    $\pm$ & 56.5 & 307.7 &  $\pm$ & 49.4 \\ 
      &      & 1.50 & 2.00& 435.2 &    $\pm$ & 53.9 & 236.9 &  $\pm$ & 41.2 \\ 
      &      & 2.00 & 2.50& 371.4 &    $\pm$ & 47.3 & 166.1 &  $\pm$ & 31.4 \\ 
      &      & 2.50 & 3.00& 313.9 &    $\pm$ & 42.0 & 140.5 &  $\pm$ & 27.3 \\ 
      &      & 3.00 & 3.50& 266.2 &    $\pm$ & 36.9 & 117.5 &  $\pm$ & 23.3 \\ 
      &      & 3.50 & 4.00& 175.3 &    $\pm$ & 27.1 & 71.1  &  $\pm$ & 16.7 \\ 
      &      & 4.00 & 5.00& 127.4 &    $\pm$ & 18.1 & 41.7  &  $\pm$ & 9.9 \\ 
      &      & 5.00 & 6.50&  63.0 &    $\pm$ & 10.3 & 20.8  &  $\pm$ & 6.1 \\ 
      &      & 6.50 & 8.00&  25.0 &     $\pm$ & 6.8 & 7.3   &  $\pm$ & 3.3 \\ 
 0.10 & 0.15 & 0.50 & 1.00& 327.5 &    $\pm$ & 62.5 & 462.9 &  $\pm$ & 85.0 \\ 
      &      & 1.00 & 1.50& 492.2 &    $\pm$ & 65.7 & 330.2 &  $\pm$ & 50.0 \\ 
      &      & 1.50 & 2.00& 317.9 &    $\pm$ & 46.3 & 235.9 &  $\pm$ & 40.6 \\ 
      &      & 2.00 & 2.50& 332.7 &    $\pm$ & 48.5 & 107.5 &  $\pm$ & 26.4 \\ 
      &      & 2.50 & 3.00& 224.1 &    $\pm$ & 36.6 & 138.4 &  $\pm$ & 28.8 \\ 
      &      & 3.00 & 3.50& 113.0 &    $\pm$ & 21.9 & 72.6  &  $\pm$ & 18.5 \\ 
      &      & 3.50 & 4.00&  97.8 &    $\pm$ & 21.7 & 57.4  &  $\pm$ & 15.3 \\ 
      &      & 4.00 & 5.00&  52.2 &    $\pm$ & 10.9 & 24.1  &  $\pm$ & 7.6 \\ 
      &      & 5.00 & 6.50&  15.1 &     $\pm$ & 5.4 & 3.7   &  $\pm$ & 2.1 \\ 
      &      & 6.50 & 8.00&   2.2 &     $\pm$ & 1.1 & 1.4   &  $\pm$ & 1.2 \\ 
 0.15 & 0.20 & 0.50 & 1.00& 337.5 &    $\pm$ & 69.6 & 244.2 &  $\pm$ & 56.7 \\ 
      &      & 1.00 & 1.50& 313.5 &    $\pm$ & 52.3 & 198.7 &  $\pm$ & 42.9 \\ 
      &      & 1.50 & 2.00& 225.6 &    $\pm$ & 40.9 & 113.9 &  $\pm$ & 30.5 \\ 
      &      & 2.00 & 2.50& 144.9 &    $\pm$ & 30.4 & 98.7  &  $\pm$ & 28.0 \\ 
      &      & 2.50 & 3.00&  78.8 &    $\pm$ & 22.3 & 45.5  &  $\pm$ & 19.3 \\ 
      &      & 3.00 & 3.50&  55.2 &    $\pm$ & 16.9 & 15.2  &  $\pm$ & 10.0 \\ 
      &      & 3.50 & 4.00&  29.3 &    $\pm$ & 11.4 & 27.5  &  $\pm$ & 12.7 \\ 
      &      & 4.00 & 5.00&  19.8 &     $\pm$ & 8.0 & 11.0  &  $\pm$ & 6.7 \\ 
      &      & 5.00 & 6.50&   6.0 &     $\pm$ & 4.5 &    &   - &  \\ 
      &      & 6.50 & 8.00&    &    - &  &    &  - &  \\ 
 0.20 & 0.25 & 0.50 & 1.00& 235.0 &    $\pm$ & 59.7 & 186.7 &  $\pm$ & 50.2 \\ 
      &      & 1.00 & 1.50& 137.0 &    $\pm$ & 43.4 & 200.3 &  $\pm$ & 58.2 \\ 
      &      & 1.50 & 2.00& 126.2 &    $\pm$ & 42.0 & 117.5 &  $\pm$ & 38.2  \\ 
      &      & 2.00 & 2.50&  70.2 &    $\pm$ & 31.5 & 115.8 &  $\pm$ & 39.6 \\ 
      &      & 2.50 & 3.00&  37.3 &    $\pm$ & 16.1 & 60.8  &  $\pm$ & 25.7 \\ 
      &      & 3.00 & 3.50&  36.8 &    $\pm$ & 16.7 & 22.1  &  $\pm$ & 15.0 \\ 
      &      & 3.50 & 4.00&  18.0 &    $\pm$ & 13.2 & 15.1  &  $\pm$ & 15.3 \\ 
      &      & 4.00 & 5.00&  11.0 &     $\pm$ & 9.2 &  4.5  &   $\pm$ & 5.5 \\ 
      &      & 5.00 & 6.50&   2.8 &     $\pm$ & 4.8 &    &   - &  \\ 
      &      & 6.50 & 8.00&    &     - &  &    &   - &  \\ 
\hline
\end{tabular}

  }
  }
\end{table}

The kinematic range of the measurements covers the
momentum region from 0.5~{\GeVc} to 8.0~{\GeVc} (subdivided into 10 intervals) and the angular
range from 0.05~{rad} to 0.25~{rad} (subdivided into 4 intervals).
The error bars correspond to
the combined statistical and systematic
errors as described in section~\ref{errorest}.
The overall normalization error of 2\% for the normalization of
incident protons and of 1\% for the target size variation are not shown.

\begin{figure}[tb]
\centering
\includegraphics[width=.49\textwidth]{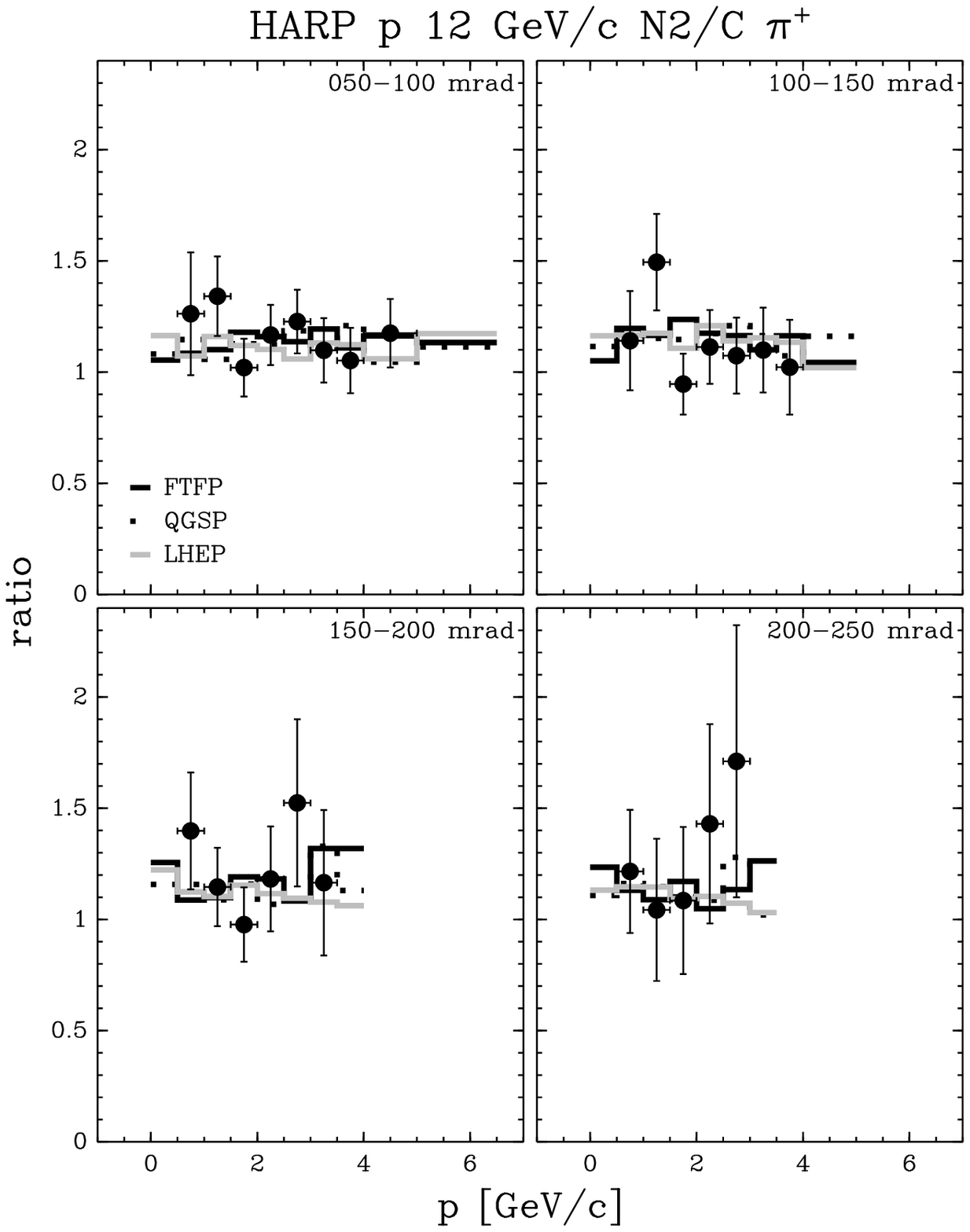}
\includegraphics[width=.49\textwidth]{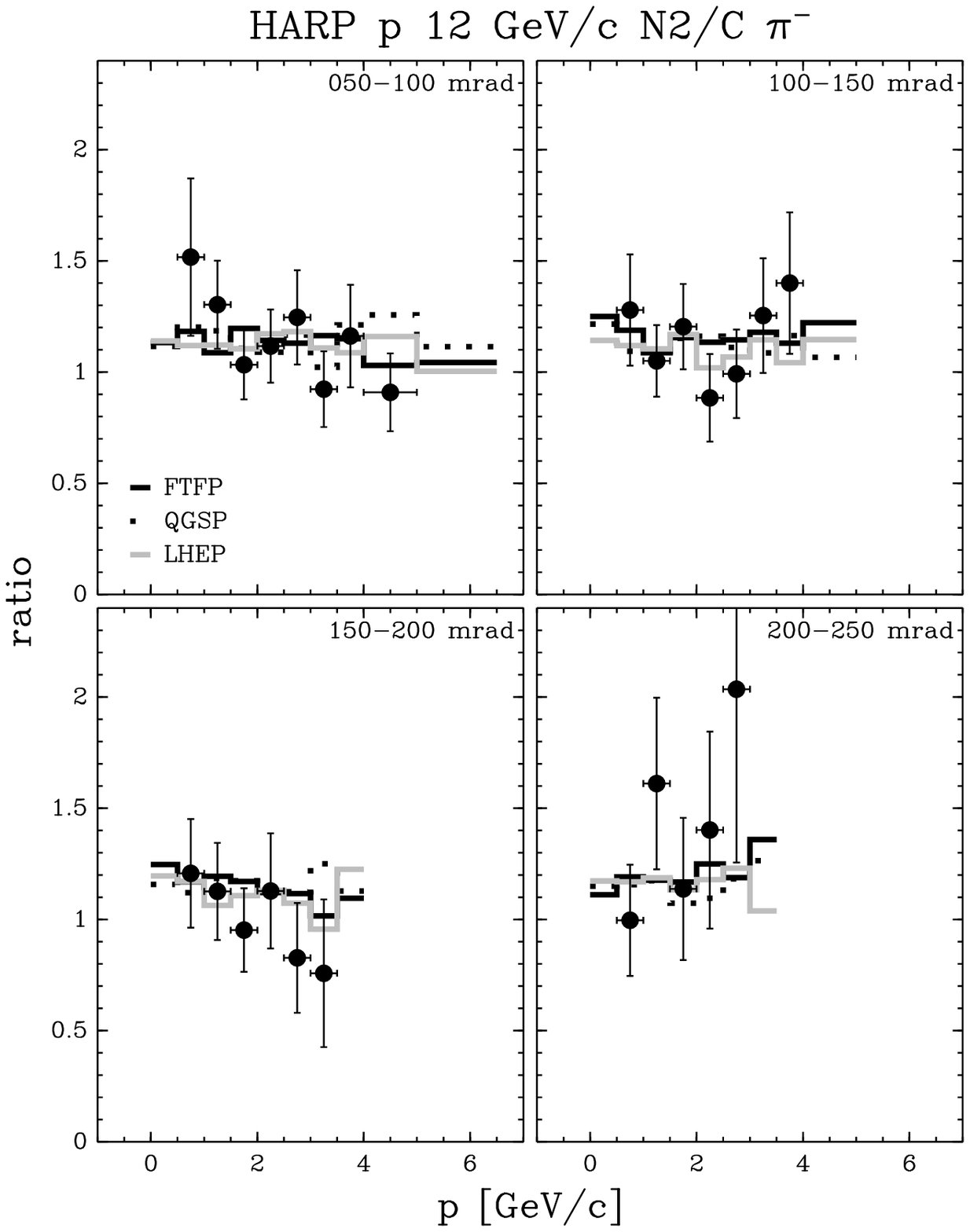}
\caption{
  p--N$_2$ to p--C production 
  ratio for $\pi^+$(left panel) and $\pi^-$(right panel)
  at 12~\GeVc, compared with GEANT4 simulation predictions
  using different models. 
In the top right corner of each plot the 
covered angular range is shown in mrad.
Only statistical errors are used, since most systematic ones cancel.}
\label{fig:ratio1}
\end{figure}
\begin{figure}[tb]
\centering
\includegraphics[width=.49\textwidth]{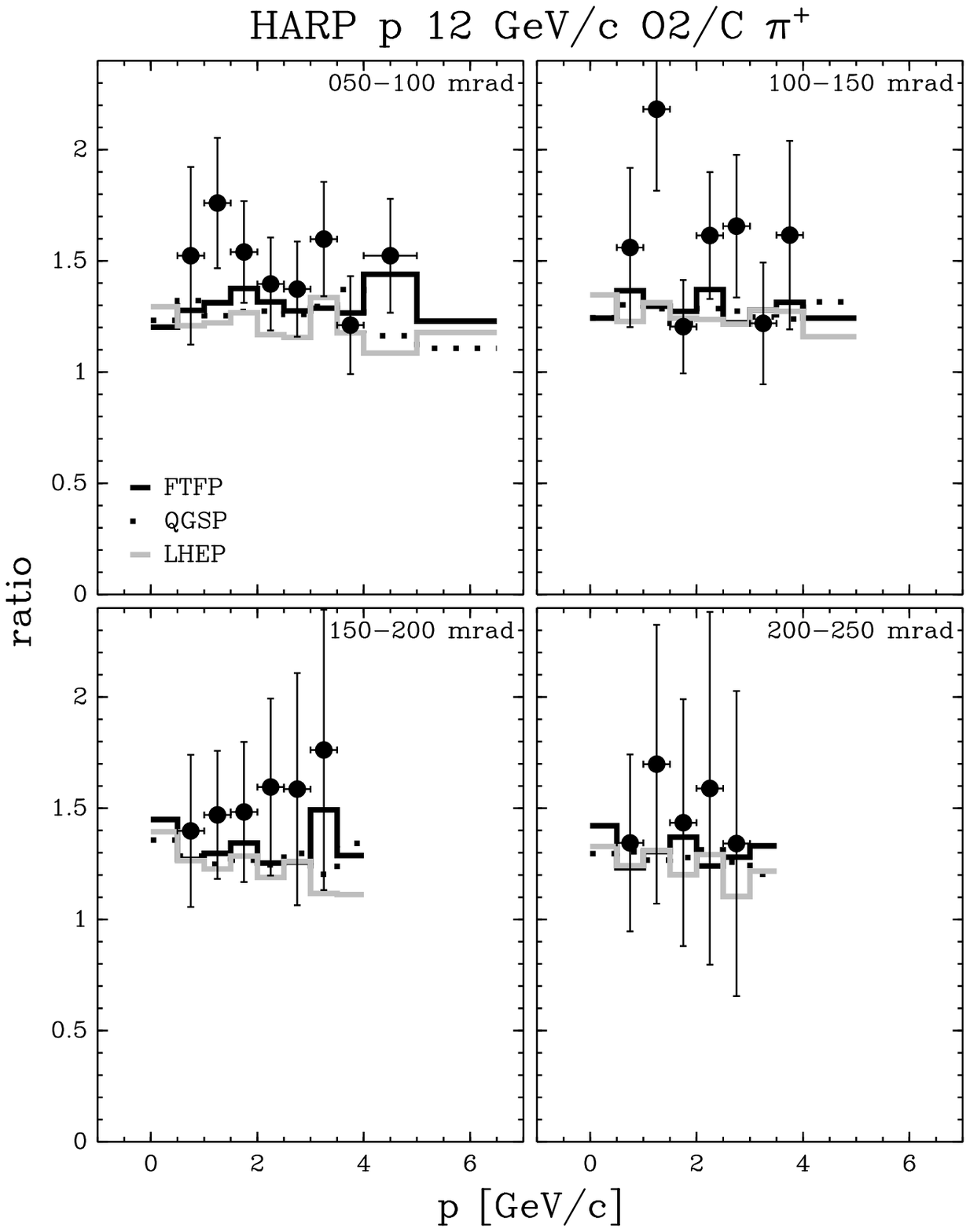}
\includegraphics[width=.49\textwidth]{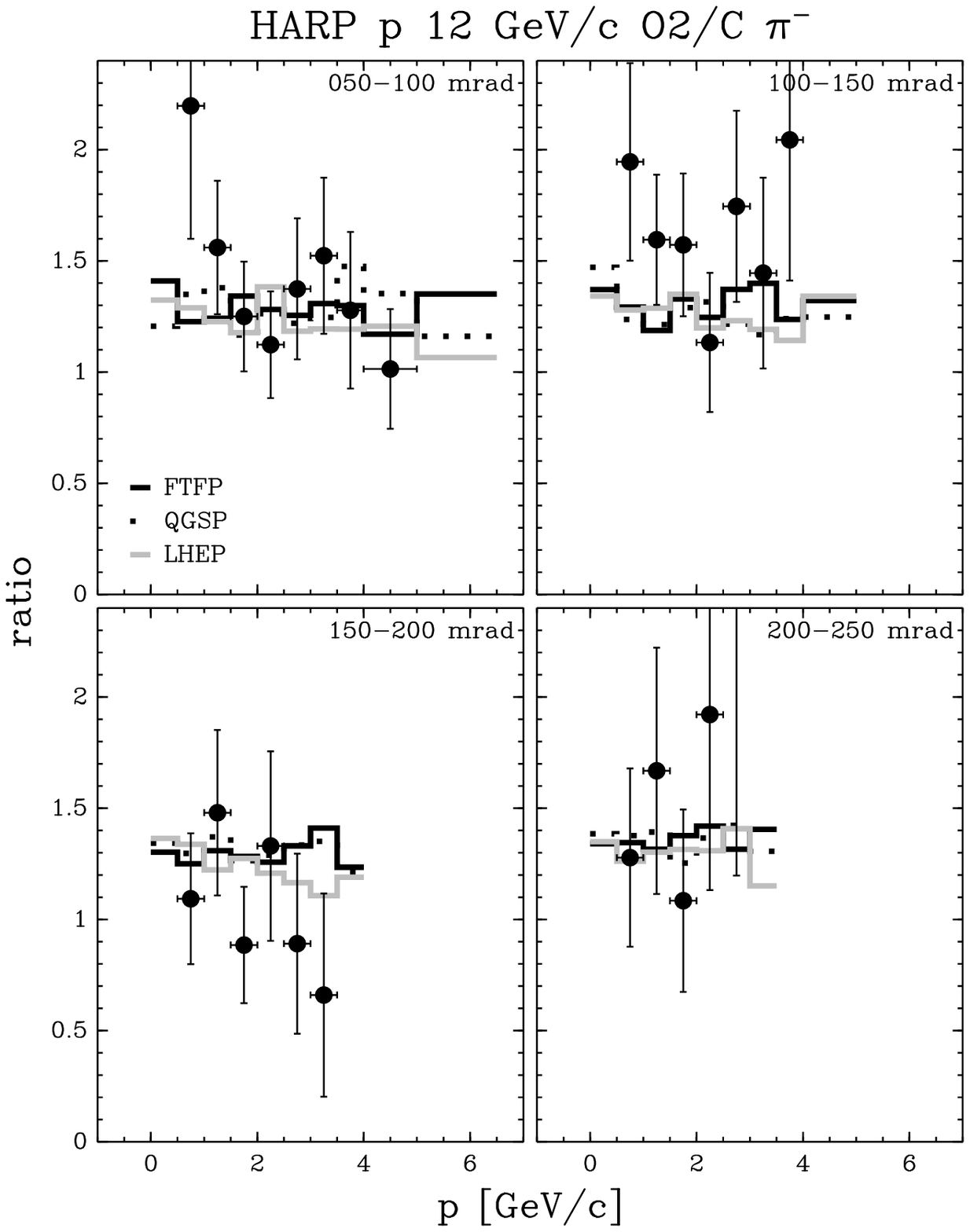}
\caption{
  p--O$_2$ to p--C production 
  ratio for $\pi^+$(left panel) and $\pi^-$(right panel)
  at 12~\GeVc, compared with GEANT4 simulation predictions
  using different models. 
In the top right corner of each plot the 
covered angular range is shown in mrad.
Only statistical errors are used, since most systematic ones cancel.}
\label{fig:ratio2}
\end{figure}

The pion production ratios N$_2$/C and O$_2$/C are presented in 
Figs.~\ref{fig:ratio1}-\ref{fig:ratio2} and are
compared to GEANT4 Monte Carlo predictions. 
As noted before, the difference between the target materials is 
justified by an overall normalisation factor taking into account the 
different nuclear masses of the target materials.
The various models (see \cite{ref:Gmod} for details) 
do predict the ratio of cross-sections accurately,
with very little spread between them.
This conclusion is different when the absolute predictions models are
compared with the measured cross-sections as shown in
Ref.~\cite{ref:carbonfw}.

\section{Summary and conclusions}\label{sec:conclusions}

The results reported in this article may  contribute to the precise 
calculations of atmospheric neutrino fluxes and to the improvement 
of our understanding of extensive air showers simulations and 
hadronic interactions at low energies. 
A detailed study of the role of hadronic interactions
for production of muons in extensive air showers, which are one
of the main ingredients to infer the mass and the energy of the
primary cosmic ray particle, is shown in reference \cite{ref:eas}.

In this paper we presented measurements of the double-differential 
production cross-section of positive and negative
pions in the collisions of 12~\GeVc protons 
with thin nitrogen and oxygen targets. 
The data were reported in bins of pion momentum and angle 
in the kinematic range 0.5~\GeVc$\leq p_\pi < 8$~\GeVc 
and 0.05~rad $\leq \theta_\pi <$ 0.25~rad in the laboratory frame. 
A detailed error analysis has been performed yielding 
total bin-to-bin differential errors (statistical and systematic) 
of about 15\%,
an overall normalization error of 2\% and additional 1\% for the target size variation.
We should stress that the HARP data are the first 
measurements with cryogenic targets in this kinematic region
with good precision.

Simulations predict that collisions of protons with a carbon target are 
very similar to proton interactions with air (see e.g. \cite{ref:christine_phd}). 
That explains  why these datasets can be used for tuning models needed in
astroparticle physics simulations. 
Our measurements on p--N$_2$ and p--O$_2$ confirm these 
predictions.

\section{Acknowledgments}

We gratefully acknowledge the help and support of the PS beam staff
and of the numerous technical collaborators who contributed to the
detector design, construction, commissioning and operation.  
In particular, we would like to thank
G.~Barichello,
R.~Brocard,
K.~Burin,
V.~Carassiti,
F.~Chignoli,
D.~Conventi,
G.~Decreuse,
M.~Delattre,
C.~Detraz,  
A.~Domeniconi,
M.~Dwuznik,   
F.~Evangelisti,
B.~Friend,
A.~Iaciofano,
I.~Krasin, 
D.~Lacroix,
J.-C.~Legrand,
M.~Lobello, 
M.~Lollo,
J.~Loquet,
F.~Marinilli,
R.~Mazza,
J.~Mulon,
L.~Musa,
R.~Nicholson,
A.~Pepato,
P.~Petev, 
X.~Pons,
I.~Rusinov,
M.~Scandurra,
E.~Usenko,
R.~van der Vlugt,
for their support in the construction of the detector
and P. Dini for his contribution to MonteCarlo production. 
The collaboration acknowledges the major contributions and advice of
M.~Baldo-Ceolin, 
L.~Linssen, 
M.T.~Muciaccia and A. Pullia
during the construction of the experiment.
The collaboration is indebted to 
V.~Ableev,
F.~Bergsma,
P.~Binko,
E.~Boter,
M.~Calvi, 
C.~Cavion,
M.~Chizov, 
A.~Chukanov,
A.~DeSanto, 
A.~DeMin, 
M.~Doucet,
D.~D\"{u}llmann,
V.~Ermilova, 
W.~Flegel,
Y.~Hayato,
A.~Ichikawa,
O.~Klimov,
T.~Kobayashi,
D.~Kustov, 
M.~Laveder, 
M.~Mass,
H.~Meinhard,
A.~Menegolli, 
T.~Nakaya,
K.~Nishikawa,
M.~Paganoni,
F.~Paleari,
M.~Pasquali,
M.~Placentino,
V.~Serdiouk,
S.~Simone,
P.J.~Soler,
S.~Troquereau,
S.~Ueda,
A.~Valassi and
R.~Veenhof
for their contributions to the experiment.

We acknowledge the contributions of 
V.~Ammosov,
G.~Chelkov,
D.~Dedovich,
F.~Dydak,
M.~Gostkin,
A.~Guskov,
D.~Khartchenko,
V.~Koreshev,
Z.~Kroumchtein,
I.~Nefedov,
A.~Semak,
J.~Wotschack,
V.~Zaets and
A.~Zhemchugov
to the work described in this paper.

 The experiment was made possible by grants from
the Institut Interuniversitaire des Sciences Nucl\'eair\-es and the
Interuniversitair Instituut voor Kernwetenschappen (Belgium), 
Ministerio de Educacion y Ciencia, Grant FPA2003-06921-c02-02 and
Generalitat Valenciana, grant GV00-054-1,
CERN (Geneva, Switzerland), 
the German Bundesministerium f\"ur Bildung und Forschung (Germany), 
the Istituto Na\-zio\-na\-le di Fisica Nucleare (Italy), 
INR RAS (Moscow) and the Particle Physics and Astronomy Research Council (UK).
We gratefully acknowledge their support.

\clearpage 
\begin{appendix}
\section{Cross-section data}\label{app:data}
The tabulated cross-section data 
for p--C interactions at 12~\GeVc, 
already published in \cite{ref:carbonfw}, are  reported again here
with a different binning for comparison.
\begin{table}[!ht]
  \small{
  \caption{\label{tab:xsec_results_pC12}
    HARP results for the double-differential $\pi^+$ and $\pi^-$ production
    cross-section in the laboratory system,
    $d^2\sigma^{\pi}/(dpd\Omega)$, for p--C interactions at 12~\GeVc. 
    Each row refers to a
    different $(p_{\hbox{\small min}} \le p<p_{\hbox{\small max}},
    \theta_{\hbox{\small min}} \le \theta<\theta_{\hbox{\small max}})$ bin,
    where $p$ and $\theta$ are the pion momentum and polar angle, respectively.
    The central value as well as the square-root of the diagonal elements
    of the covariance matrix are given.}
  \centerline{
    \begin{tabular}{|c|c|c|c|rcr|rcr|} \hline
$\theta_{\hbox{\small min}}$ &
$\theta_{\hbox{\small max}}$ &
$p_{\hbox{\small min}}$ &
$p_{\hbox{\small max}}$ &
\multicolumn{3}{c|}{$d^2\sigma^{\pi^+}/(dpd\Omega)$} &
\multicolumn{3}{c|}{$d^2\sigma^{\pi^-}/(dpd\Omega)$} 
\\
(rad) & (rad) & (GeV/c) & (GeV/c) &
\multicolumn{3}{c|}{(mb/(GeV/c sr))} &
\multicolumn{3}{c|}{(mb/(GeV/c sr))}
\\ \hline

 0.05 & 0.10 & 0.50 & 1.00& 190.6 & $\pm$  &27.1& 132.3 &   $\pm$ &21.5 \\ 
      &      & 1.00 & 1.50& 237.3 &   $\pm$  & 23.0& 197.2 &  $\pm$ &20.9 \\ 
      &      & 1.50 & 2.00& 282.6 &  $\pm$   &23.2& 189.6 &  $\pm$ &17.9 \\ 
      &      & 2.00 & 2.50& 266.1 &    $\pm$ &21.1& 147.9 &  $\pm$ &14.8 \\ 
      &      & 2.50 & 3.00& 228.6 &    $\pm$ &18.1& 102.2 &  $\pm$ &12.7 \\ 
      &      & 3.00 & 3.50& 166.6 &   $\pm$  &13.6& 77.2  &  $\pm$ &9.1  \\ 
      &      & 3.50 & 4.00& 144.7 &   $\pm$  &13.7& 55.7  &  $\pm$ &8.1  \\ 
      &      & 4.00 & 5.00&  83.6 &    $\pm$ & 7.5& 41.2  &  $\pm$ &5.0  \\ 
      &      & 5.00 & 6.50&  36.5 &    $\pm$  &4.2& 8.6   &  $\pm$ &2.3  \\ 
      &      & 6.50 & 8.00&  16.4 &     $\pm$ &2.5& 1.6   &  $\pm$ &0.9  \\ 
 0.10 & 0.15 & 0.50 & 1.00& 209.9 &    $\pm$ &26.9& 238.0 &  $\pm$ &32.3 \\ 
      &      & 1.00 & 1.50& 225.6 &    $\pm$ &23.1& 207.0 &  $\pm$ &21.4 \\ 
      &      & 1.50 & 2.00& 264.1 &    $\pm$ &25.3& 150.1 &  $\pm$ &16.5 \\ 
      &      & 2.00 & 2.50& 206.1 &    $\pm$ &20.4& 94.9  &  $\pm$ &11.9 \\ 
      &      & 2.50 & 3.00& 135.4 &    $\pm$ &14.2& 79.3  &  $\pm$ &10.4 \\ 
      &      & 3.00 & 3.50&  92.7 &    $\pm$ &10.5& 50.2  &  $\pm$ &7.7  \\ 
      &      & 3.50 & 4.00&  60.5 &     $\pm$ &8.4& 28.1  &  $\pm$ &4.4  \\ 
      &      & 4.00 & 5.00&  37.3 &    $\pm$  &4.7& 17.2  &  $\pm$ &3.7  \\ 
      &      & 5.00 & 6.50&   9.5 &     $\pm$ &1.7& 2.6   &  $\pm$ &0.9  \\ 
      &      & 6.50 & 8.00&   2.6 &     $\pm$ &0.7& 0.2   &  $\pm$ &0.1  \\ 
 0.15 & 0.20 & 0.50 & 1.00& 241.4 &   $\pm$  &31.7& 223.5 &  $\pm$ &30.5 \\ 
      &      & 1.00 & 1.50& 213.2 &    $\pm$ &21.9& 134.2 &  $\pm$ &17.2 \\ 
      &      & 1.50 & 2.00& 152.1 &    $\pm$ &16.8& 128.7 &  $\pm$ &16.1 \\ 
      &      & 2.00 & 2.50&  90.8 &    $\pm$ &12.3& 74.2  &  $\pm$ &11.0 \\ 
      &      & 2.50 & 3.00&  49.7 &     $\pm$& 8.4& 51.1  &  $\pm$ &8.5  \\ 
      &      & 3.00 & 3.50&  31.3 &     $\pm$ &5.8& 23.0  &  $\pm$ &4.8  \\ 
      &      & 3.50 & 4.00&  24.4 &     $\pm$ &5.2& 11.3  &  $\pm$ &3.0  \\ 
      &      & 4.00 & 5.00&  11.3 &     $\pm$ &2.6& 6.7   &  $\pm$ &2.0  \\ 
      &      & 5.00 & 6.50&   3.7 &     $\pm$ &1.4& 0.5   &  $\pm$ &0.4  \\ 
      &      & 6.50 & 8.00&   0.6 &     $\pm$ &0.3&    &  - & \\ 
 0.20 & 0.25 & 0.50 & 1.00& 174.8 &    $\pm$ &26.6& 146.1 &  $\pm$ &23.5 \\ 
      &      & 1.00 & 1.50&  80.7 &    $\pm$ &15.4& 120.1 &  $\pm$ &19.4 \\ 
      &      & 1.50 & 2.00&  87.9 &    $\pm$ &17.3& 108.4 &  $\pm$ &21.0 \\ 
      &      & 2.00 & 2.50&  44.2 &     $\pm$& 9.6& 60.2  &  $\pm$ &13.7 \\ 
      &      & 2.50 & 3.00&  27.8 &     $\pm$ &7.7& 24.5  &  $\pm$ &7.3  \\ 
      &      & 3.00 & 3.50&  18.0 &     $\pm$ &5.4&  10.4 &  $\pm$ &3.8  \\ 
      &      & 3.50 & 4.00&   8.1 &     $\pm$ &3.5&  4.9  &  $\pm$ &2.5 \\ 
      &      & 4.00 & 5.00&   5.1 &     $\pm$ &3.1&  1.1  &  $\pm$ &1.3  \\ 
      &      & 5.00 & 6.50&   1.9 &     $\pm$ &1.4&    &  - &  \\ 
      &      & 6.50 & 8.00&    &     - & &  & - &   \\
\hline
\end{tabular} 

  }
  }
\end{table}

\end{appendix}

\end{document}